\documentclass[12pt]{article}
 \usepackage{amsmath}    
 \usepackage{amssymb}
 \usepackage{url}
 \usepackage{graphicx}   
\usepackage{textcomp}
\usepackage{ifxetex}
\ifxetex
    \usepackage[cm-default,no-math]{fontspec} 
    \usepackage{xltxtra} 
    \usepackage{polyglossia} 
    \setdefaultlanguage{russian} 
    \setotherlanguage{english} 
    \defaultfontfeatures{Mapping=tex-text} 
    \newfontfamily\russianfont{Charis SIL}
    \newfontfamily\englishfont{Charis SIL}
\else
    \usepackage[T1,T2A]{fontenc}
    \usepackage[utf8]{inputenc}
    \usepackage[russian,english]{babel}
\fi
\usepackage{color,graphicx,shortvrb}
\usepackage{url}
\usepackage{wrapfig}
\usepackage{wasysym}
\usepackage[bookmarks, bookmarksnumbered, bookmarksopen,colorlinks=true,linkcolor=black]{hyperref}

\usepackage{hyphenat}
  
\begin{document}
%
%
%

%
\selectlanguage{english}
\begin{center}
\textbf{\Large On the Theory of Wave Packets}\\[0.5cm]
\textbf{\Large D.~V.~Naumov\\[0.5cm]}
\end{center}
\begin{tabular}{l}
\textit{\indent JINR, 141980, Dubna, Russia}\\
\textit{\indent e-mail: dnaumov@jinr.ru}
\end{tabular}

\date{\today}

\begin{abstract}
In this paper we discuss some aspects of the theory of wave packets. We consider a popular non-covariant 
Gaussian model used in various applications and show that it predicts too slow a longitudinal dispersion rate for
relativistic particles. We revise this approach by considering a covariant model of Gaussian wave packets, and examine our results by inspecting a wave packet of arbitrary form.  A general formula for the time
dependence of the dispersion of a wave packet of arbitrary form is found. Finally, we give a transparent interpretation of the disappearance of the wave function over time due to the dispersion --- a feature often considered undesirable, but which is unavoidable for wave packets. We find, starting from simple examples, proceeding with their generalizations and finally by considering the continuity equation, that the  integral over time of both the flux and probability densities are asymptotically proportional to the factor $1/|\mathbf{x}|^2$ in the rest frame of the wave packet, just as in the case of an ensemble of classical particles.
\end{abstract}

\newpage
\section{Introduction}
A classical or quantum object composed of multiple waves with trajectory characteristics of a solid body 
is often called a {\em wave packet}. A prime example of a quantum wave packet is the wave function describing the free propagation of a particle. Waves in fluids and gases are examples of classical wave packets in our everyday life. 

Wave packets are known to spread with the passage of time. In other words the spatial size of the wave packet grows over time while their amplitude vanishes. This well known unavoidable feature of wave packets is often considered their ``disadvantage'' in various applications. 

In particular while it is impossible to build a consistent scattering theory without wave packets~\cite{Taylor:1972,Goldberger:1964} it is not rare to meet an argument that their vanishing with time makes them not quite adequate objects for initial and final states. Indeed, 
the initial (final) state is defined at infinitely far past (future) moment in time at which any wave packet should vanish thus making their use in $S$-theory sometimes arguable. 

However, as we show in this paper this problem is not critical for the $S$-matrix formalism, as the spreading of wave packets has a very clear interpretation.  The dispersion leads precisely to a $1/|\mathbf{x}|^2$ suppression of the time-integrated probability to observe the particle, just as one would expect for an ensemble of classical ``particles'' thus leading to transparent normalization factors of the initial and final states.

We will first introduce some definitions and notations used within this paper, and then outline its layout. 
Throughout the paper we will use the Natural units where $\hbar=c=1$. For definiteness we will examine a quantum wave packet, while our results will also be valid for classical wave packets because we do not consider particle creation or annihilation in this paper. The wave packet for a spinless particle with mass $m$ has the form:
\begin{equation}
 \label{eq:WP_state_1}
 |\text{wave packet}\rangle = \int \frac{d\mathbf{k}\; \phi(\mathbf{k})}{(2\pi)^3 2E_\mathbf{k}}|\mathbf{k}\rangle,
\end{equation}
where $|\mathbf{k}\rangle$ is the Fock state with definite 3-momentum $\mathbf{k}$ and energy $E_\mathbf{k} =
\sqrt{\mathbf{k}^2+m^2}$, and $\phi(\mathbf{k})$ is a Lorentz-invariant function assumed to be narrow around some
momentum $\mathbf{p}$ which we will define later on. As we will show in Sec.~\ref{sec:generalWP} a real-valued
$\phi(\mathbf{k})$ corresponds to the wave packet with mean three-dimensional position: 
\[
\langle\mathbf{x}\rangle=0 \text{ at time }  t=0.
\] 
The action of the translation operator  $\text{e}^{i\hat{P}x_0}$ (where $\hat{P}$ is the
operator of 4-momentum, and $x_0=(t_0,\mathbf{x}_0)$ the 4-dimensional displacement vector) onto the wave packet state
in \eqref{eq:WP_state_1}:
\begin{equation}
  \text{e}^{i\hat{P}x_0}\int \frac{d\mathbf{k}\; \phi(\mathbf{k})}{(2\pi)^3 2E_\mathbf{k}}|\mathbf{k}\rangle =  
 \int\frac{d\mathbf{k}\;\text{e}^{i kx_0}\phi(\mathbf{k})}{(2\pi)^3 2E_\mathbf{k}}|\mathbf{k}\rangle,
\end{equation}
translates the wave packet state in space and time. Therefore, the wave packet in \eqref{eq:WP_state_1} with  
\begin{equation}
\label{eq:WP_state_2}
\phi(\mathbf{k})=\text{e}^{i kx_0}|\phi(\mathbf{k})| 
\end{equation}
has the mean position $\langle\mathbf{x}\rangle=\mathbf{x}_0$ at $t=t_0$. Since it essentially gives us no new
information, in this paper we will consider the case where $\mathbf{x}_0=0$ at $t_0=0$, assuming a real-valued
$\phi(\mathbf{k})$ function.

In coordinate space the wave packet is characterized  by the Lorentz-invariant wave function, which can be
obtained by projecting the wave packet state ~\eqref{eq:WP_state_1} 
onto the $\langle 0|\hat{\Psi}(x)$ state, where $\hat{\Psi}(x)$ is the quantum field operator for the (pseudo)scalar 
particle, as follows:
\begin{equation}
 \label{eq:psi_definition}
  \psi(x) = \langle 0|\hat{\Psi}(x)|\text{wave packet}\rangle = \int
\frac{d\mathbf{k}\;\phi(\mathbf{k})}{(2\pi)^3 2E_\mathbf{k}}\text{e}^{-i kx}. 
\end{equation}
It is apparent that $\psi(x)$ satisfies the Klein-Gordon equation. If $\psi(0,\mathbf{x})$ is known then $\phi(\mathbf{k})$
could be found as follows:
\begin{equation}
 \label{eq:phi_from_psi}
 \frac{\phi(\mathbf{k})}{2E_\mathbf{k}} = \int
d\mathbf{x}\;\psi(0,\mathbf{x})\text{e}^{-i\mathbf{k}\mathbf{x}}.
\end{equation}
The 4-vector of the flux density $j_\mu(x) = (\rho(x), \mathbf{j}(x))$ is defined in the usual way for the Klein-Gordon
equation:
\begin{equation}
 \label{eq:4flux}
 j_\mu(x)  = i\left(\psi^*(x)\partial_\mu\psi(x) - \psi(x)\partial_\mu\psi^*(x)\right).
\end{equation}
The  probability density, which is not relativistic invariant, can be normalized in a relativistically invariant way as follows:
\begin{equation}
 \label{eq:rho_integral_3D}
  \int d\mathbf{x}\rho(t,\mathbf{x}) = \int \frac{d\mathbf{k}\left|\phi(\mathbf{k})\right|^2}{(2\pi)^3
2E_{\mathbf{k}}}
= 1
\end{equation}
which corresponds to one particle within the Universe. The spatial integral over the flux density $\mathbf{j}(t,\mathbf{x})$ is equal to the mean velocity of the wave packet, as can be seen from the following:
\begin{equation}
 \label{eq:wp_velocity}
 \int d\mathbf{x}\;\mathbf{j}(t,\mathbf{x}) = \int \frac{d\mathbf{k}\left|\phi(\mathbf{k})\right|^2}{(2\pi)^3
2E_{\mathbf{k}}}\mathbf{v}_{\mathbf{k}} = \langle\mathbf{v}\rangle.
\end{equation}
By definition, the mean energy $\langle \text{E}\rangle$ and mean momentum $\langle\mathbf{P}\rangle$ of the wave
packet state are obtained from:
\begin{equation}
 \langle \text{P}_\mu\rangle  = \frac{\langle\text{wave packet}|\hat{\text{P}}_\mu|\text{wave
packet}\rangle}{\langle\text{wave packet}|\text{wave packet}\rangle},
\end{equation}
where $\hat{\text{P}}_\mu$ is the $\mu$-th component of the 4-momentum operator acting on the Fock state as
$\hat{\text{P}}_\mu|\mathbf{k}\rangle=k_\mu|\mathbf{k}\rangle$. Therefore:
\begin{align}
 \label{eq:mean_energy}
  \langle \text{E}\rangle   = \int\frac{d\mathbf{k}|\phi(\mathbf{k})|^2}{(2\pi)^32E_\mathbf{k}}E_\mathbf{k}\\
 \label{eq:mean_momentum}
  \langle \mathbf{P}\rangle = \int\frac{d\mathbf{k}|\phi(\mathbf{k})|^2}{(2\pi)^32E_\mathbf{k}}\mathbf{k}
\end{align}
Eqs.~\eqref{eq:WP_state_1},~\eqref{eq:rho_integral_3D},~\eqref{eq:wp_velocity},~\eqref{eq:mean_energy},
~\eqref{eq:mean_momentum} suggest that
$|\phi(\mathbf{k})|^2/2E_\mathbf{k}$ is the probability density to have 3-momentum $\mathbf{k}$ for the state in
~\eqref{eq:WP_state_1}.
The paper is organized as follows. We begin with a well-known example of a non-covariant Gaussian wave packet in
Sec.~\ref{sec:noncovariantWP} to illustrate the main features of a dispersive wave packet. We observe that the wave
packet disperses in both the longitudinal and transverse directions relative to the mean velocity vector. However, it is shown that the speed of the longitudinal dispersion predicted by the non-covariant model is too slow after examining the covariant Gaussian model of the wave packet given in Sec.~\ref{sec:covariantWP}. For the considered examples we will find that the time-integrated flux density asymptotically follows $1/|\mathbf{x}|^2$. In
Sec.~\ref{sec:generalWP} we generalize our calculations for wave packets of arbitrary form restricted by
~\eqref{eq:WP_state_2}. We will find that, on average, a wave packet follows the classical trajectory. We will also produce a general formula for the time dependence of the dispersion of the wave packet. After inspecting this formula we realize that the prediction of the non-covariant Gaussian model for the longitudinal dispersion is indeed not correct for ultra-relativistic particles, while the covariant model agrees with the prediction using more general considerations. In Sec.~\ref{sec:flux_asymptotics_general} we examine the asymptotic natures of the time-integrated flux and probability densities, and confirm that a wave packet of arbitrary form will follow a $1/|\mathbf{x}|^2$ trend. In Sec.~\ref{sec:continuity} we revisit the $1/|\mathbf{x}|^2$ tendencies from another point of view --- by considering possible conclusions from the continuity equation, which holds true not only for the Klein-Gordon equation, from which we performed all of our calculations in this paper, but also for other equations like 
those of Schroedinger and Dirac. Finally, in Sec.~\ref{sec:conclusion} we discuss the main results of this paper and draw conclusions.

\section{Gaussian Wave Packet}
\subsection{\label{sec:noncovariantWP}Non-Covariant Gaussian Wave Packet}
Let us consider, as an useful illustration, a well-known example of a non-covariant Gaussian wave packet with
$\phi(\mathbf{k}) = \varphi_{\rm G}(\mathbf{k})$ assumed to be:
\begin{equation}
 \label{eq:phi_non_covariant}
 \varphi_{\rm G}(\mathbf{k}) =
\sqrt{2E_\mathbf{p}}\left(\frac{2\pi}{\sigma_{\rm p}^2}\right)^{3/4}\text{e}^{-\frac{\left(\mathbf{k}-\mathbf{p}
\right)^2 } { 4\sigma_{\rm p} ^2}},
\end{equation}
where $\sigma_{\rm p}$ is the {\em constant} Gaussian width in the momentum distribution of the wave packet. 
The space-time wave function $\psi(x)=\psi_{\rm G}(x)$ could be obtained from ~\eqref{eq:psi_definition} by assuming small enough $\sigma_{\rm p}$ to expand $E_\mathbf{k}$ in the exponent around $\mathbf{k}=\mathbf{p}$:
\begin{equation}
 \label{eq:Ek_expansion}
 E_\mathbf{k} = E_{\mathbf{p}} + \mathbf{v}_{\mathbf{p}}(\mathbf{k}-\mathbf{p}) +
\frac{m^2}{2E_{\mathbf{p}}^3}(\mathbf{k}-\mathbf{p})^2 +
\frac{\left(\mathbf{p}\times\mathbf{k}\right)^2}{2E_{\mathbf{p}}^3}+\dots,
\end{equation}
where $\mathbf{v}=\mathbf{p}/E_\mathbf{p}$, and by replacing $E_\mathbf{k}$ in the denominator of
$\phi(\mathbf{k})/2E_\mathbf{k}$ with $E_\mathbf{p}$. Then the remaining Gaussian integrals in 
\eqref{eq:psi_definition} yield:
\begin{equation}
 \label{eq:psi_definition_noncovariant}
 \psi_{\rm G}(x) =
\frac{\exp{\left[-i px-\frac{\left(\mathbf{x}_{\rm L}-\mathbf{v}t\right)^2}{
4\sigma_{\rm x}^2(1+i t/\tau_{\rm L})}
-\frac{\mathbf{x}_{\rm T}^2}{4\sigma_{\rm x}^2(1+i t/\tau_{\rm T})}\right]}}{(2\pi)^{3/4}\sqrt{2E_\mathbf{p}}
\sigma_{\rm x}^{3/2}\sqrt { (1+i t/\tau_{\rm L})} (1+i t/\tau_{\rm T})} 
\end{equation}
where  $p=(E_\mathbf{p},\mathbf{p})$, $\sigma_{\rm x}^2 = 1/4\sigma_{\rm p}^2$ and 
\begin{equation}
 \label{eq:tau_non_covariant}
  \tau_{\rm L}  = \gamma_\mathbf{p}^3\tau\quad,
  \tau_{\rm T} = \gamma_\mathbf{p}\tau, \quad 
  \tau=2\sigma_{\rm x}^2m, \quad \gamma_\mathbf{p} = \frac{E_\mathbf{p}}{m}.
\end{equation}
$\mathbf{x}_{\rm L}$ and $\mathbf{x}_{\rm T}$ are components of $\mathbf{x}$ parallel and perpendicular, respectively, to the average (and most probable) velocity vector $\mathbf{v}$.

As one might observe, $\psi_{\rm G}(x)$ describes a wave packet which spreads over time. To present it in a more
transparent fashion let us examine $|\psi_{\rm G}(x)|$:
\begin{equation}
 \label{eq:psi_definition_noncovariant_mod}
 |\psi_{\rm G}(t,\mathbf{x})| = 
\frac{\exp{\left[-\frac{\left(\mathbf{x}_{\rm L}-\mathbf{v}t\right)^2}{
4\sigma_{\rm x}^2(1+t^2/\tau_{\rm L}^2)}
-\frac{\mathbf{x}_{\rm T}^2}{4\sigma_{\rm x}^2(1+t^2/\tau_{\rm T}^2)}\right]}}{(2\pi)^{3/4}\sqrt{2E_\mathbf{p}}
\sigma_{\rm x}^{
3/2}\left(1+t^2/\tau_{\rm L}^2\right)^{1/4}\sqrt{1+t^2/\tau_{\rm T}^2}} 
\end{equation}
Considering this example, a non-covariant Gaussian wave packet is described at $t=0$ by:
\begin{equation}
 \label{eq:psi_definition_noncovariant_t0}
 \psi_{\rm G}(0,\mathbf{x}) = \frac{1}{(2\pi)^{3/4}\sqrt{2E_\mathbf{p}}\sigma_{\rm x}^{
3/2}}
\exp{
\left[
i\mathbf{p}\mathbf{x}-\frac{\mathbf{x}^2}{4\sigma_{\rm x}^2}
\right]}
\end{equation}
and at later times spreads in both the longitudinal and transverse directions. The squares of the longitudinal ($\sigma^2_{\rm xL}(t)$) and
transverse ($\sigma^2_{\rm xT}(t)$) dispersions read:
\begin{align}
 \label{eq:dispersions_non_covariant}
 \sigma^2_{\rm xL}(t)  & = \sigma^2_{\rm x}(1+t^2/\tau_{\rm L}^2),\\
 \sigma^2_{\rm xT}(t)  & = \sigma^2_{\rm x}(1+t^2/\tau_{\rm T}^2),
\end{align}
where $\tau_{\rm L}$ and $\tau_{\rm T}$, given by \eqref{eq:tau_non_covariant}, are related to each other by
$\tau_{\rm L} = \gamma^2_\mathbf{p}\tau_{\rm T}$. In what follows we will refer to  $\tau_{\rm L}$ and
$\tau_{\rm T}$ as the longitudinal and transverse dispersion times, respectively. $\tau$ is the dispersion in the rest frame of the wave packet when, obviously, $\tau_L=\tau_T=\tau$.

The non-covariant model suggests that the wave packet is expected to disperse more slowly in the longitudinal direction by a factor of $\gamma^2_\mathbf{p}$ as compared to the transverse direction. There are two dispersion regimes: transverse dispersion ($t\gg\tau_{\rm T}$) and longitudinal dispersion ($t\gg
\tau_{\rm L}$). In the regime of complete dispersion one obtains:
\begin{align}
 \label{eq:dispersions_non_covariant_L_2}
 \sigma_{\rm xL}(t)  & = \sigma_{\rm x}\frac{t}{\tau}\frac{1}{\gamma^3_\mathbf{p}}, \quad  t\gg \tau_{\rm L}\\
 \label{eq:dispersions_non_covariant_T_2}
 \sigma_{\rm xT}(t)  & = \sigma_{\rm x}\frac{t}{\tau}\frac{1}{\gamma_\mathbf{p}},  \quad t\gg\tau_{\rm T}
\end{align}

It should come as no surprise, however, that the non-covariant wave packet in ~\eqref{eq:phi_non_covariant} might fail in predicting relativistic effects. Indeed, as we will demonstrate in Sec.~\ref{sec:covariantWP}, a relativistically covariant version of the Gaussian wave packet predicts an alternate dependence on the longitudinal
dispersion. Therefore, we will postpone a qualitative and quantitative discussion of dispersion effects until
Sec.~\ref{sec:covariantWP}.

Let us now see how these dispersions lead to a $1/|\mathbf{x}|^2$ suppression of the time-integrated flux density.
For simplicity we will perform the calculations in the rest frame of the wave packet. The flux density reads:
\begin{align}
 \mathbf{j}_{\rm G}(t,\mathbf{x})  & = -i\left(\psi_{\rm G}^*(\mathbf{x},t)\nabla \psi_{\rm G}(\mathbf{x},t) -
\psi_{\rm G}(\mathbf{x},t)\nabla
\psi_{\rm G}^*(\mathbf{x},t)\right)\nonumber\\
&  =
\frac{\mathbf{x}\;t/\tau\exp{\left[-\frac{\mathbf{x}^2}{
2\sigma_{\rm x}^2(1+t^2/\tau^2)}\right]}}{(2\pi)^{3/2}2m\sigma_{\rm x}^{
5} \left(1+t^2/\tau^2\right)^{5/2}}
\label{eq:flux_non_covariant}
\end{align}
An explicit calculation of the time integral $\int_0^\infty d t \;\mathbf{j}_{\rm G}(t,\mathbf{x})$ allows us to observe
that, in the regime $|\mathbf{x}|^2\gg \sigma_{\rm x}^2$, one has the following transparent formula:
\begin{equation}
 \label{eq:Phi_non_covariant}
 \boldsymbol\Phi_{\rm G}(\mathbf{x}) = \int_0^\infty d t \;\mathbf{j}_{\rm G}(t,\mathbf{x}) =
\frac{\mathbf{x}}{4\pi|\mathbf{x}|^3}.
\end{equation}
The corrections to ~\eqref{eq:Phi_non_covariant} are suppressed by $\text{e}^{-\mathbf{x}^2/2\sigma_{\rm x}^2}$.

\subsection{\label{sec:covariantWP}Covariant Gaussian Wave Packet}
Let us note that relativistically covariant wave packet does not neccessarily imply near speed of light velocity of the latter.  A relativistically covariant Gaussian wave packet  was considered in ~\cite{Naumov:2010um,Naumov:2009zza}. The proposed $\phi(\mathbf{k})= \phi_{\rm RG}(\mathbf{k})$ function is explicitly Lorentz-invariant and reads as follows:
\begin{equation}
 \label{eq:phi_covariant_1}
 \phi_{\rm RG}(\mathbf{k}) = N_{\rm RG}\exp{\left[\frac{(p-k)^2}{4\sigma_{\rm p}^2}\right]}.
\end{equation}
While an exact formula for the corresponding space-time wave function $\psi(x)=\psi_{\rm RG}(x)$ was obtained in
~\cite{Naumov:2010um,Naumov:2009zza}, we proceed here in an approximate fashion in order to more quickly attain the main results.
Let us again exploit the smallness of $\sigma^2_{\rm p}$ and use ~\eqref{eq:Ek_expansion} to produce:
\begin{equation}
 \label{eq:phi_covariant_2}
 \phi_{\rm RG}(\mathbf{k}) =
\sqrt{2m}\left(\frac{2\pi}{\sigma_{\rm p}^2}\right)^{3/4}\!\!\!\!\!\exp{\left[-\frac{(\mathbf{p}-\mathbf{k}_{\text{L}
} )^2
} { 4\sigma_{\rm pL}^2} -
\frac{\mathbf{k}_{\rm T}^2}{4\sigma_{\rm pT}^2}\right]},
\end{equation}
where 
\begin{equation}
  \sigma_{\rm pL}^2 = \sigma_{\rm p}^2\gamma^2_\mathbf{p}, \quad
  \sigma_{\rm pT}^2 = \sigma_{\rm p}^2
\end{equation}
and the relativistically invariant normalization constant was obtained with help of ~\eqref{eq:rho_integral_3D}.
The coordinate wave function $\psi_{\rm RG}(x)$ can be obtained in the same way as we proceeded for
~\eqref{eq:psi_definition_noncovariant}:
\begin{equation}
 \label{eq:psi_definition_covariant}
 \psi_{\rm RG}(x) =
\frac{\exp{\left[-i px-\frac{\left(\mathbf{x}_{\rm L}-\mathbf{v}t\right)^2}{
4\sigma_{\rm xL}^2(1+i t/\tau_\mathbf{p})}
-\frac{\mathbf{x}_{\rm T}^2}{4\sigma_{\rm xT}^2(1+i t/\tau_\mathbf{p})}\right]}}{(2\pi)^{3/4}
\sqrt {2m} \sigma^3_{\rm x}(1+i t/\tau_\mathbf{p})^{3/2}}, 
\end{equation}
where 
\begin{align}
 \label{eq:tau_covariant}
 \sigma_{\rm xL}^2  = \frac{1}{4\sigma_{\rm pL}^2} = \frac{\sigma_{\rm x}^2}{\gamma^2_\mathbf{p}}, \quad  
 \sigma_{\rm xT}^2 \quad = \frac{1}{4\sigma_{\rm pT}^2} =
 \sigma_{\rm x}^2, \quad \tau_\mathbf{p} = \tau \gamma_\mathbf{p}
\end{align}
with $\sigma_{\rm x}^2 = 1/4\sigma_\mathbf{p}^2$ and $\tau=2m\sigma_{\rm x}^2$, just as in the case of the 
non-covariant wave packet.
Similarly to Sec.~\ref{sec:noncovariantWP} one can obtain the longitudinal and transverse dispersions as functions of
time:
\begin{align}
 \label{eq:dispersions_covariant}
 \sigma^2_{\rm xL}(t)  & = \sigma^2_{\rm xL}(1+t^2/\tau^2_\mathbf{p}),\\
 \sigma^2_{\rm xT}(t)  & = \sigma^2_{\rm xT}(1+t^2/\tau^2_\mathbf{p}),
\end{align}
where $\sigma^2_{\rm xL}$ and $\sigma^2_{\rm xT}$ are given by ~\eqref{eq:tau_covariant}.  Now in the regime of
complete dispersion ($t\gg\tau$) one has:
\begin{align}
 \label{eq:dispersions_covariant_L_2}
 \sigma_{\rm xL}(t)  & = \sigma_{\rm xL}(0)\frac{t}{\tau_\mathbf{p}}, \\
 \label{eq:dispersions_covariant_T_2}
 \sigma_{\rm xT}(t)  & = \sigma_{\rm xT}(0)\frac{t}{\tau_\mathbf{p}}, \\
 \label{eq:dispersions_covariant_LT_2}
 \sigma_{\rm xL}(t)  & = \frac{1}{\gamma_\mathbf{p}}\sigma_{\rm xT}(t).
\end{align}
Comparing~\eqref{eq:dispersions_covariant_L_2},~\eqref{eq:dispersions_covariant_T_2},~\eqref{eq:dispersions_covariant_LT_2} to ~\eqref{eq:dispersions_non_covariant_L_2},~\eqref{eq:dispersions_non_covariant_T_2} one might observe a
problem with the non-covariant model's $\varphi_{\rm G}(\mathbf{k})$ (see ~\eqref{eq:phi_non_covariant}), which
wrongly predicts, by the factor $1/\gamma^2_\mathbf{p}$, too slow a dispersion in the longitudinal direction as compared to the covariant model $\phi_{\rm RG}(\mathbf{k})$ (see ~\eqref{eq:phi_covariant_1}) where both longitudinal and transverse dispersions have the same rate. However, since  the size of the longitudinal spatial width is smaller than the corresponding transverse width by a factor of $\gamma_\mathbf{p}$, the absolute value of the former always remains smaller. 

In order to give some quantitative estimates of both the longitudinal and transverse dispersions let's consider three examples: an electron, a neutrino with a mass of $0.1$ eV, and a ``classical`` particle with a mass of 1 gram. Let us assume that initially these ``particles`` had in their rest frames $\sigma_{\rm x}=1\mu m$ and see how long it will take to double the corresponding longitudinal and transverse sizes.  We will consider two cases as an example: the particle is at rest, or has a full energy equal to 1 GeV. We summarize $\tau_{\rm L}$ and $\tau_{\rm T}$ in Tab.~\ref{tab:taus} for non-covariant and $\tau$ for covariant Gaussian models (see
~\eqref{eq:dispersions_non_covariant_L_2},~\eqref{eq:dispersions_non_covariant_T_2} and ~\eqref{eq:dispersions_covariant_L_2},~\eqref{eq:dispersions_covariant_T_2},~\eqref{eq:dispersions_covariant_LT_2}.
\begin{table}[htb]
\label{tab:taus}
\begin{center}
\begin{tabular}{c|c|c|c|c}
\hline
      &  & \multicolumn{2}{c|}{non covariant}            & \multicolumn{1}{c}{covariant}\\
\hline
mass      & $\gamma_\mathbf{p}$      & $\tau_{\rm L}$    & $\tau_{\rm T}$ & $\tau_\mathbf{p}$ \\
\hline
$0.5\cdot 10^{6}$ eV & $1$& $5\cdot 10^{-8}\text{s}$ & $5\cdot 10^{-8}\text{s}$ & $5\cdot 10^{-8}$\text{s}\\
$0.5\cdot 10^{6}$ eV & $2\cdot 10^{3}$& $4\cdot 10^{2}\text{s}$ & $10^{-4}\text{s}$ &$10^{-4}$\text{s}\\
$0.1$ eV & $1$& $10^{-14}\text{s}$ & $10^{-14}\text{s}$ &$10^{-14}\text{s}$\\
$0.1$ eV & $10^{10}$& $10^{16}\text{s}$ & $10^{-4}\text{s}$ & $10^{-4}$\text{s}\\
1 g  & $1$& $3\cdot 10^{11}\text{y}$ & $3\cdot 10^{11}\text{y}$ &$3\cdot 10^{11}\text{y}$\\
\hline
\end{tabular}
\caption{
$\tau_{\rm L}$ and $\tau_{\rm T}$  for particles with masses $0.5$ MeV, $0.1$ eV and a ``classical`` particle
with a mass of 1 gram. It is assumed that the particles are either at rest or have a total energy of 1 GeV. 
Estimates are given for both the non-covariant and covariant Gaussian models.}
\end{center}
\end{table}

As one can see from this table the coordinate wave functions $\psi(x)$ of particles with microscopic masses
quickly disperse in the rest frames of the particles. However, the predictions of the non-covariant and covariant
Gaussian models are sharply different for the longitudinal dispersion rates. Essentially, the non-covariant model produces too slow a longitudinal dispersion (by a factor $\gamma^2_\mathbf{p}$) compared to the covariant model.
This makes a dramatic difference. For example, a particle with neutrino mass on the order of $0.1$ eV does not 
disperse longitudinally during the lifetime of the Universe according to the non-covariant model, while it 
disperses quite quickly (during $10^{-4}$ seconds) according to the covariant model.

Let us note also that the coordinate wave functions $\psi(x)$ of a ``particle`` with a mass on the order of 1 gram,
initially bound within a space of $1\mu m$, disperses within times significantly exceeding the lifetime of the Universe, thus bridging quantum and classical physics.

It is worth noting that the relativistic wave packet model ~\eqref{eq:phi_covariant_1} also displays similar behavior at $\mathbf{x}^2\gg \sigma^2_{\rm x}$ as in ~\eqref{eq:Phi_non_covariant}. As we will show in the next section this is a general property of wave packets of arbitrary form.

\section{\label{sec:generalWP}Wave Packet of an Arbitrary Form}
In this section it will be beneficial to use the following representation of $j_\mu(x)$, which can be
obtained with help of \eqref{eq:psi_definition}:
\begin{align}
 \label{eq:rho}
 \rho(x) = \int d\mathbf{k}d\mathbf{q}\;\Pi(\mathbf{k},\mathbf{q})\text{e}^{-i (k-q)x}, \\
 \mathbf{j}(x)   = \int d\mathbf{k}d\mathbf{q}\;\mathbf{J}(\mathbf{k},\mathbf{q})\text{e}^{-i (k-q)x}. 
\end{align}
where 	
\begin{align}
  \Pi(\mathbf{k},\mathbf{q})  = \frac{\phi(\mathbf{k})\phi(\mathbf{q})}{(2\pi)^6
2E_{\mathbf{k}}2E_{\mathbf{q}}}\left[E_{\mathbf{k}}+E_{\mathbf{q}}\right],\\
  \boldsymbol{\mathbf{J}}(\mathbf{k},\mathbf{q})  = \frac{\phi(\mathbf{k})\phi(\mathbf{q})}{(2\pi)^6
2E_{\mathbf{k}}2E_{\mathbf{q}}}\left[{\mathbf{k}}+{\mathbf{q}}\right].
\end{align}

\subsection{Trajectory and Dispersion of the Wave Packet}
It is well known that the mean coordinate of the wave packet follows the classical trajectory. Indeed, explicit calculation for an arbitrary form of real-valued $\phi(\mathbf{k})$ yields
\begin{align}
\langle\mathbf{x}\rangle  = \int  d\mathbf{x}\rho(t,\mathbf{x})\mathbf{x} = \int \frac{d\mathbf{k}\;\phi^2(\mathbf{k})}{(2\pi)^3 2E_\mathbf{k}}
\frac{\mathbf{k}}{E_\mathbf{k}}t = \langle\mathbf{v}\rangle t
\label{eq:3coordinate}
\end{align}
where $\langle\mathbf{v}\rangle$ is given by \eqref{eq:wp_velocity}. 

Let us now examine the time-dependence of the coordinate dispersion. By definition, the square of the spatial dispersion reads:
\begin{equation}
 \label{eq:dispersion_WP_1}
 \sigma^2_{\rm x}(t) = \langle\mathbf{x}^2\rangle - \langle\mathbf{x}\rangle^2.
\end{equation}
Performing calculations for $\langle\mathbf{x}^2\rangle$ yields:
\begin{align}
 \label{eq:dispersion_WP_2}
  \langle\mathbf{x}^2\rangle & = \int d\mathbf{x}\; \mathbf{x}^2\rho(t,\mathbf{x}) = t^2\int
\frac{d\mathbf{k}\;\phi^2(\mathbf{k})}{(2\pi)^32E_\mathbf{k}}\mathbf{v}_k^2-\int
\frac{d\mathbf{k}\;\phi(\mathbf{k})}{(2\pi)^32E_\mathbf{k}}\frac{\partial^2\phi(\mathbf{k})}{\partial\mathbf{k}^2}
\\
& +\int
\frac{d\mathbf{k}}{(2\pi)^3}\left[\phi^2(\mathbf{k})\left(\frac{m^2}{4E_\mathbf{k}^5}-\frac{\mathbf{v}^2_\mathbf{k}}{
2E_\mathbf{k}^3}\right)+\frac{\mathbf{v}_\mathbf{k}}{4E_\mathbf{k}^2}\frac{\partial\phi^2(\mathbf{k})}{\partial\mathbf
{k}}\right].\nonumber
\end{align}
Taking into account the integral
\begin{align}
 \label{eq:dispersion_WP_3}
  \int
\frac{d\mathbf{k}}{(2\pi)^3}\frac{\mathbf{v}_\mathbf{k}}{4E_\mathbf{k}^2}\frac{\partial\phi^2(\mathbf{k})}{
\partial\mathbf{k}} & = - \int
\frac{d\mathbf{k}\phi^2(\mathbf{k})}{(2\pi)^3}\frac{\partial}{\partial\mathbf{k}}\frac{\mathbf{v}_\mathbf{k}}{
4E_\mathbf{k}^2} \\
& = -
\int\frac{d\mathbf{k}\phi^2(\mathbf{k})}{(2\pi)^3}\left(\frac{m^2}{4E_\mathbf{k}^5}-\frac{\mathbf{v}_k^2}{2E_\mathbf{
k }
^3} \right), \nonumber
\end{align}
the last line of ~\eqref{eq:dispersion_WP_2} is precisely cancelled. The integral \[-\int
\frac{d\mathbf{k}\phi(\mathbf{k})}{(2\pi)^32E_\mathbf{k}}\frac{\partial^2\phi(\mathbf{k})}{\partial\mathbf{k}^2}\]
in ~\eqref{eq:dispersion_WP_2} does not depend on space-time coordinates.  It has the dimensions of the
3-coordinate squared and it is not invariant under the Lorentz transformations. Let us denote it by:
\begin{equation}
 \label{eq:sigma_general}
 \sigma^2_{\rm x} = -\int
\frac{d\mathbf{k}\;\phi(\mathbf{k})}{(2\pi)^32E_\mathbf{k}}\frac{\partial^2\phi(\mathbf{k})}{\partial\mathbf{k}^2}.
\end{equation}
 Therefore, 
\begin{equation}
 \label{eq:dispersion_WP_4}
  \langle\mathbf{x}^2\rangle = \sigma^2_{\rm x} + \langle\mathbf{v}^2\rangle t^2,
\end{equation}
where 
\begin{equation}
 \label{eq:meanV2}
 \langle\mathbf{v}^2\rangle = \int
\frac{d\mathbf{k}|\phi(\mathbf{k})|^2}{(2\pi)^32E_\mathbf{k}}\mathbf{v}_\mathbf{k}^2.
\end{equation}
Thus, the square of the dispersion of the coordinates reads:
\begin{align}
 \label{eq:dispersion_WP_5}
 \sigma^2_{\rm x}(t)  &  = \sigma^2_{\rm x} +
\left(\langle\mathbf{v}^2\rangle-\langle\mathbf{v}\rangle^2\right) t^2\nonumber\\
  &= \sigma^2_{\rm x} + \sigma^2_\mathbf{v} t^2.
\end{align}
where $\sigma^2_\mathbf{v} = \langle\mathbf{v}^2\rangle-\langle\mathbf{v}\rangle^2$ is the velocity dispersion. The general result \eqref{eq:dispersion_WP_5} being one of the main results of this paper is not however new. It was already obtained in \cite{Almeida:1983ep} where the derivation was unfortunatelly somewhat inconsistent as the authors asummed $|\psi(t,\mathbf{x})|^2$ to be a probability density function while $\rho(t,\mathbf{x})$ defined in \eqref{eq:4flux} instead should  be used for the scalar relativistic particle.

We will now examine how the coordinate dispersion depends on relativistic effects, keeping in mind the discussions in Sec.~\ref{sec:noncovariantWP} and Sec.~\ref{sec:covariantWP} concerning different predictions for the
longitudinal dispersion of the non-covariant and covariant Gaussian models of wave packets.

First, we rewrite $\sigma^2_\mathbf{v}$ as follows:
\begin{align}
 \label{eq:dispersion_general_1}
  \sigma^2_\mathbf{v} & = \langle\mathbf{v}^2\rangle-\langle\mathbf{v}\rangle^2 = 
\langle\mathbf{v}_{\rm T}^2\rangle+\langle\mathbf{v}_{\rm L}^2\rangle-\langle\mathbf{v}\rangle^2\nonumber\\
 & =\langle\mathbf{v}_{\rm T}^2\rangle+\langle(\mathbf{v}_{\rm L}-\langle\mathbf{v}\rangle)^2\rangle, 
\end{align}
where $\langle\mathbf{v}_{\rm T}^2\rangle$ and $\langle\mathbf{v}_{\rm L}^2\rangle$ are, respectively, the means of
the squares of the longitudinal and transverse projections of wave packet velocities relative to the mean velocity vector.
Rewriting $\langle\mathbf{v}_{\rm T}^2\rangle$ and
$\langle(\mathbf{v}_{\rm L}-\langle\mathbf{v}\rangle)^2\rangle$ using the variables in the rest frame of the
wave packet:
\begin{align}
E_\mathbf{k}         & = \gamma_{\langle\mathbf{v}\rangle}(E_\mathbf{k}^*+\langle\mathbf{v}\rangle
\mathbf{k}^*_{\rm L}), \\ 
\mathbf{k}_{\rm L}  & = \gamma_{\langle\mathbf{v}\rangle}(\mathbf{k}^*_{\rm L}+\langle\mathbf{v}\rangle
E_{\mathbf{k}^*}), \quad \mathbf{k}_{\rm T} = \mathbf{k}_{\rm T}^*,\\
\mathbf{v}_{\mathbf{k}{\rm L}}  & =
\frac{\mathbf{v}^*_{\mathbf{k}{\rm
L}}+\langle\mathbf{v}\rangle}{1+\mathbf{v}^*_{\mathbf{k}{\rm L}}\langle\mathbf{v } 
\rangle}, \quad \gamma_{\langle\mathbf{v}\rangle} = \frac{1}{\sqrt{1-\langle\mathbf{v}\rangle^2}}
\end{align}
as follows:
\begin{align}
 \label{eq:vt2_general}
 \langle\mathbf{v}_{\rm T}^2\rangle  
 & = \int\frac{d\mathbf{k}|\phi(\mathbf{k})|^2}{(2\pi)^32E_\mathbf{k}}\mathbf{v}_{\mathbf{k}{\rm T}}^2 \nonumber\\
 & =\frac{1}{\gamma^2_{\langle\mathbf{v}\rangle}}
\int\frac{d\mathbf{k}^*|\phi(\mathbf{k^*})|^2}{(2\pi)^32E_{\mathbf{k}^*}}\frac{\mathbf{v}^{*2}_{\mathbf{k}{\rm T}}}{
(1+\langle\mathbf{v}\rangle\mathbf{v^{*}}_{\mathbf{k}{\rm L}})^2}
\end{align}
and 
\begin{align}
 \label{eq:vl2_general}
\langle(\mathbf{v}_{\rm L}-\langle\mathbf{v}\rangle)^2\rangle
& =\int\frac{d\mathbf{k}|\phi(\mathbf{k})|^2}{(2\pi)^32E_\mathbf{k}}(\mathbf{v}_{\mathbf{k}{\rm L}}-\langle\mathbf{v}
\rangle)^2 \\
 & =\frac{1}{\gamma^4_{\langle\mathbf{v}\rangle}}
\int\frac{d\mathbf{k}^*|\phi(\mathbf{k^*})|^2}{(2\pi)^32E_{\mathbf{k}^*}}\frac{\mathbf{v}^{*2}_{\mathbf{k}{\rm L}}}{
(1+\langle\mathbf{v}\rangle\mathbf{v^*}_{\mathbf{k}{\rm L}})^2}\nonumber
\end{align}
Comparing ~\eqref{eq:vt2_general} to ~\eqref{eq:vl2_general} one can observe that for narrow wave packets one
gets, to the first order:
\begin{align}
 \langle\mathbf{v}_{\rm T}^2\rangle
 & =\frac{1}{\gamma^2_{\langle\mathbf{v}\rangle}}\langle\mathbf{v}^{*2}_{\rm T}\rangle\\ 
 \langle(\mathbf{v}_{\rm L}-\langle\mathbf{v}\rangle)^2\rangle
 & =\frac{1}{\gamma^4_{\langle\mathbf{v}\rangle}}\langle\mathbf{v}^{*2}_{\rm L}\rangle 
\end{align}
Therefore, in the regime of complete dispersion, keeping in mind that in the rest frame of the wave packet $\langle\mathbf{v}^{*2}_{\rm L}\rangle =
\langle\mathbf{v}^{*2}_{\rm T}\rangle/2 = \frac{1}{3}\langle\mathbf{v}^{*2}\rangle$ and using
~\eqref{eq:dispersion_WP_5}, one obtains:
\begin{align}
  \sigma^2_{\rm xL}(t) & = \frac{1}{3\gamma^4_{\langle\mathbf{v}\rangle}}\langle\mathbf{v}^{*2}\rangle t^2,\\
  \sigma^2_{\rm xT}(t) & = \frac{2}{3\gamma^2_{\langle\mathbf{v}\rangle}}\langle\mathbf{v}^{*2}\rangle t^2,\\
  \sigma^2_{\rm xL}(t) & = \frac{1}{2\gamma^2_{\langle\mathbf{v}\rangle}}\sigma^2_{\rm xT}(t)
 \label{eq:dispersion_WP_6}
\end{align}
in agreement with calculations performed for the covariant model (see
~\eqref{eq:dispersions_covariant_L_2},~\eqref{eq:dispersions_covariant_T_2},~\eqref{eq:dispersions_covariant_LT_2}).
The seemingly extra factor of $1/2$ in ~\eqref{eq:dispersion_WP_6} is due to the cumulative nature of the definition in
~\eqref{eq:dispersion_WP_1}, which adds together all projections of the dispersion.  Apparently, in the rest frame of
the wave packet: 
\[
 \sigma^2_{\rm x}(t) = \sigma^2_{\rm xL}(t) + \sigma^2_{\rm xT}(t) =\langle\mathbf{v}^{*2}\rangle t^2.
\]

Let's examine the following question: If the Gaussian wave packet can be shown to exhibit an asymptotic behavior of
$\boldsymbol{\Phi}(\mathbf{x}) = \mathbf{x}/|\mathbf{x}|^3$ (see ~\eqref{eq:Phi_non_covariant}), does this also hold 
true in the general case?
As we will see in Sec.~\ref{sec:flux_asymptotics_general} this is indeed true. Moreover, keeping in mind that probability  $\rho(x)$
and flux $\mathbf{j}(x)$ densities are closely related to each other, it implies that a similar asymptotic behavior should
also apply to the integral $\int_0^\infty d t \rho(t,\mathbf{x})$. Indeed, for the wave function with definite
4-momentum  ($\psi(x)=Ne^{-i px}$), the relation is obvious:
\begin{equation}
  \label{eq:rho_flux}
  \rho(x) = 2E_\mathbf{p}|N|^2, \quad \mathbf{j}(x) = 2\mathbf{p}|N|^2, \quad \mathbf{j}(x) = \mathbf{v}\rho(x).
\end{equation}
Apparently, for narrow wave packets, a relation similar to ~\eqref{eq:rho_flux} should be valid as well. In
Sec.~\ref{sec:rho_asymptotics_general} we will explicitly calculate the asymptotic behavior of the
time-integrated probability densities.

\subsection{\label{sec:flux_asymptotics_general}Asymptotic Behavior of the Time-Integrated Flux Density}
Let us compute here
\begin{equation}
\boldsymbol{\Phi}(\mathbf{x}) \equiv \int_0^\infty d t \; \mathbf{j}(t,\mathbf{x}). 
 \label{eq:Phi_general}
\end{equation}

In order to integrate over the time in ~\eqref{eq:Phi_general} let us use the following formula:
\begin{equation}
\int_0^\infty d t \text{e}^{\pm i\alpha t}   = \pi\left(\delta(\alpha) \pm
\frac{i}{\pi}\mathcal{P}\frac{1}{\alpha}\right)
 \label{eq:singular_exp}
\end{equation}
where $\mathcal{P}$ represents the Cauchy principal value of the integral.

Therefore, 
\begin{equation}
\boldsymbol{\Phi}(\mathbf{x}) = \boldsymbol{\Phi}_1(\mathbf{x})+\boldsymbol{\Phi}_2(\mathbf{x}) 
\end{equation}
where 
\begin{align}
  {\boldsymbol{\Phi}}_1(\mathbf{x})  \!&=\! \pi\!\int\!\!
d\mathbf{k}d\mathbf{q}\;\mathbf{J}(\mathbf{k},\mathbf{q})\delta(E_\mathbf{k} -E_\mathbf { q } )\cos\left [
(\mathbf
{k}\!-\!\mathbf{q})\mathbf{x}\right]\\
  \boldsymbol{\Phi}_2(\mathbf{x})  \!&=\! \int\!\!
d\mathbf{k}d\mathbf{q}\;\mathbf{J}(\mathbf{k},\mathbf{q})\mathcal{P}\left(\frac{1}{E_\mathbf{k}-E_\mathbf{q}}
\right)\sin\left[(\mathbf{k}\!-\!\mathbf{q})\mathbf{x}\right]
 \label{eq:Phi_12_general}
\end{align}
It is easy to see that in the rest frame of the wave packet $\boldsymbol{\Phi}_1(\mathbf{x})=0$, because changing the
integration variables $\mathbf{k}\to-\mathbf{k}$ and $\mathbf{q}\to-\mathbf{q}$ changes the sign of the integrand
$\mathbf{J}(\mathbf{k},\mathbf{q})\to \mathbf{J}(\mathbf{-k},\mathbf{-q}) = -
\mathbf{J}(\mathbf{k},\mathbf{q})$, and in the rest frame of the wave packet $\phi(\mathbf{k})$ depends only on
the absolute value of $\mathbf{k}$ and not on its direction.

For the remaining integrals in ~\eqref{eq:Phi_12_general} one could work out the angular integrations in
\begin{align}
  {\boldsymbol{\Phi}}_2(\mathbf{x})   & = \mathcal{P}\int_0^\infty d\text{k} d\text{q}  \int d\mathbf{n}_1
d\mathbf{n}_2\frac{\text{k}^2 \text{q}^2
\phi(\text{k}\mathbf{n}_1)\phi(\text{q}\mathbf{n}_2)}{(2\pi)^6
2E_{\rm k}2E_{\rm q}}\times\nonumber\\
 &\times\left[\text{k}\mathbf{n}_1+\text{q}\mathbf{n}_2\right]
\frac{1}{E_{\rm k}-E_{\rm q}}\sin\left[(\text{k}\mathbf{n}_1-\text{q}\mathbf{n}_2)\mathbf{x}\right]
 \label{eq:Phi2_general_2}
\end{align}
by again keeping in mind that $\phi(\text{k}\mathbf{n})$ does not depend on the direction vector $\mathbf{n}$ in the rest frame of
the wave packet and using the following: 
\begin{align}
 &  \int d\mathbf{n}_1
d\mathbf{n}_2\left[\text{k}\mathbf{n}_1+\text{q}\mathbf{n}_2\right]\sin\left[(\text{k}\mathbf{n}_1-\text{q}\mathbf{n}
_2)\mathbf {x} \right ] = -\frac{(4\pi)^2}{2\text{k}\text{q}}\frac{\mathbf{x}}{|\mathbf{x}|^3}\nonumber\\        
      &  \left[(\text{k}-\text{q})\sin ((\text{k}+\text{q})|\mathbf{x}|) -
(\text{k}+\text{q})\sin((\text{k}-\text{q})|\mathbf{x}|)
\right].
 \label{eq:angular_integrals}
\end{align}
Using ~\eqref{eq:angular_integrals} and an obvious identity:
\[
 \frac{1}{E_{\rm k}-E_{\rm q}} = \frac{E_{\rm k}+E_{\rm q}}{(\text{k}-\text{q})(\text{k}+\text{q})}
\]
~\eqref{eq:Phi2_general_2} can be written in the following way: 
\begin{align}
  \boldsymbol{\Phi}_2(\mathbf{x}) 
& =-\frac{1}{32\pi^4}\frac{\mathbf{x}}{|\mathbf{x}|^3}\mathcal{P}\int_0^\infty d\text{k} d\text{q}  \frac{\text{k}
\text{q}(E_{\rm k}+E_{\rm q})\phi(\text{k})\phi(\text{q})}{E_{\rm k}E_{\rm q}}\times\nonumber\\           
& \times\left[\frac{\sin
((\text{k}+\text{q})|\mathbf{x}|)}{\text{k}+\text{q}} - \frac{\sin((\text{k}-\text{q})|\mathbf{x}|)}{\text{k}-\text{q}}
\right]
 \label{eq:Phi2_general_3}
\end{align}
One might notice that  the singularities introduced by a generalized function
$\mathcal{P}(E_\mathbf{k}-E_\mathbf{q})^{-1}$  due to ~\eqref{eq:singular_exp} disappear thanks to the corresponding
$\sin((\text{k}\pm\text{q})|\mathbf{x}|)$ in the numerator of the integrals in ~\eqref{eq:Phi2_general_3}. While the
remaining integrals in  ~\eqref{eq:Phi2_general_3} can be calculated only for explicit forms of the function $\phi(\text{k})$,
we could proceed further by noting that, in the limit $|\mathbf{x}|\to\infty$, the terms  
\[
 \frac{\sin((\text{k}\pm\text{q})|\mathbf{x}|)}{\text{k}\pm\text{q}}
\]
could be replaced by delta functions with the argument $\text{k}\pm\text{q}$ due to the following relation:
\begin{equation}
 \lim_{x\to \infty} \frac{\sin \alpha x}{\alpha} = \pi\delta(\alpha).
\end{equation}
The first delta function $\delta(\text{k}+\text{q})$ cancels the integral because the integrand is zero at
$\text{q}=\text{k}=0$. A non-zero contribution comes from the second delta function $\delta (\text{k}-\text{q})$.
Therefore:                                                              
\begin{equation}
 \boldsymbol{\Phi}_2(\mathbf{x})=\frac{\mathbf{x}}{16\pi^3|\mathbf{x}|^3}\int_0^\infty d\text{k}
\frac{\text{k}^2\phi^2(\text{k})}{E_{\rm k}} =\frac{1}{4\pi}\frac{\mathbf{x}}{|\mathbf{x}|^3}
\label{eq:Phi2_general_4} 
\end{equation}
where one might notice that the remaining integral in ~\eqref{eq:Phi2_general_4} reduces to the normalization integral in
~\eqref{eq:rho_integral_3D} (multiplied by $4\pi^2$) in the rest frame of the wave packet.

To obtain ~\eqref{eq:Phi2_general_4} we used the limit $|\mathbf{x}|\to\infty$. At what distance $|\mathbf{x}|$ will
~\eqref{eq:Phi2_general_4} be a good approximation? A dimensional analysis suggests that if
$\phi(\text{k})$ can be characterized by a certain momentum ``width`` $\sigma_{\rm p}$ then this approximation works
with a good accuracy for $|\mathbf{x}|\gg 1/\sigma_{\rm p}$. Therefore, we have just proved that, for a wave packet of
arbitrary form, the time-integrated flux density displays the asymptotic behavior:
\begin{equation}
 \boldsymbol{\Phi}(\mathbf{x}) = \frac{1}{4\pi}\frac{\mathbf{x}}{|\mathbf{x}|^3} \text{ at }
|\sigma_{\rm p}\mathbf{x}|\gg 1
\end{equation}

\subsection{\label{sec:rho_asymptotics_general}Asymptotic Behavior of the Time-Integrated Probability Density}
We will now inspect the asymptotic nature of the time-integrated probability density:
\begin{equation}
 P(\mathbf{x})\equiv \int_0^\infty d t\;\rho(\mathbf{t,\mathbf{x}}) = P_1(\mathbf{x})+P_2(\mathbf{x}),
 \label{eq:P_general_1}
\end{equation}
where $P_{1,2}(\mathbf{x})$ are defined via ~\eqref{eq:singular_exp} as follows:
\begin{align}
 P_1(\mathbf{x})  &= \int
d\mathbf{k}d\mathbf{q}\;\Pi(\mathbf{k},\mathbf{q})\pi\delta(E_\mathbf{k}-E_\mathbf{q})\cos\left [
(\mathbf{k}-\mathbf{q})\mathbf{x}\right]\\
 P_2(\mathbf{x})  &= \int
d\mathbf{k}d\mathbf{q}\;\Pi(\mathbf{k},\mathbf{q})\mathcal{P}\frac{1}{E_\mathbf{k}-E_\mathbf{q}}\sin\left[
(\mathbf{k}-\mathbf{q})\mathbf{x}\right]
\end{align}
In the rest frame of the wave packet $P_2(\mathbf{x})=0$. As is easy to see, changing the variables $\mathbf{k}\to
-\mathbf{k}$ and $\mathbf{q}\to -\mathbf{q}$ changes the sign of the integrand while the integration limits remain the
same. Let us compute $P_1(\mathbf{x})$.

\begin{align}
 \label{eq:P_1}
 P_1(\mathbf{x})  & =\frac{1}{|\mathbf{x}|^2}\int_0^\infty \frac{d\text{k} \text{k}\phi^2(\text{k})}{(2\pi)^3}\sin^2
(\text{k}|\mathbf{x}|)\nonumber\\  
&=\frac{1}{4\pi|\mathbf{x}|^2}\int \frac{d \mathbf{k}\phi^2(\mathbf{k})}{(2\pi)^3
2E_\mathbf{k}}\frac{E_\mathbf{k}}{\text{k}} - \delta P_1(\mathbf{x})
\end{align}
where 
\begin{equation}
 \delta P_1(\mathbf{x}) = 
\frac{1}{|\mathbf{x}|^2}\frac{\partial}{\partial |\mathbf{x}|}\int_0^\infty \frac{d\text{k}
\text{k}\phi^2(\text{k})}{(2\pi)^3}\frac{\sin(2\text{k}|\mathbf{x}|)}{2\text{k}}
\end{equation}
One may note that
\[
 \lim_{|\mathbf{x}|\to\infty}\delta P_1(\mathbf{x}) = 0
\]
because in this limit 
\[
\lim_{|\mathbf{x}|\to\infty}\frac{\sin 2\text{k}|\mathbf{x}|}{2k} = \pi \delta(2\text{k})
\]
and 
\[
 \lim_{|\mathbf{x}|\to\infty}\int_0^\infty \frac{d\text{k}
\text{k}\phi^2(\text{k})}{(2\pi)^3}\frac{\sin(2\text{k}|\mathbf{x}|)}{2\text{k}} =
\frac{\text{k}\phi^2(\text{k})}{2(2\pi)^3}\left|_{\text{k}=0} = 0.\right.
\]
Therefore:
\begin{equation}
 \label{eq:P_2}
 P(\mathbf{x})=\frac{1}{4\pi|\mathbf{x}|^2} \left\langle|\mathbf{v}|^{-1}\right\rangle \text{ at } 
|\sigma_{\rm p}\mathbf{x}|\gg 1.
\end{equation}
Thus we have shown that the time-integrated probability density also scales as $1/|\mathbf{x}|^2$ with a coefficient
of proportionality equal to the mean of the inverse absolute value of the wave packet velocity, as was suggested by
~\eqref{eq:rho_flux}:
\begin{equation}
  \label{eq:rho_flux_asym}
  \int_0^\infty d t\; \rho(t,\mathbf{x}) = \left\langle|\mathbf{v}|^{-1}\right\rangle \int_0^\infty
d t\;|\mathbf{j}(t,\mathbf{x})|
\end{equation}
Let us note, however, that despite the fact that the mean velocity in the rest frame of the wave
packet is zero, its mean absolute value (and therefore its inverse) is not zero. This is in contrast to the
plain wave, where the velocity in the rest frame of a particle is zero, making it impossible for ~\eqref{eq:rho_flux_asym}
to be the plain wave solution.

As we have shown in this section, the asymptotic behavior of $1/|\mathbf{x}|^2$ is valid for both the time-integrated probability
density $\rho(t,\mathbf{x})$ and the flux density $\mathbf{j}(t,\mathbf{x})$ for wave packets of arbitrary form
satisfying the Klein-Gordon equation. We will now show that the $1/|\mathbf{x}|^2$ behavior applies to more general situations
following from the continuity equation, which also holds true for solutions to both the Schroedinger and Dirac equations.

\section{\label{sec:continuity}The Continuity Equation and $1/|\mathbf{x}|^2$}
The continuity equation for a quantum state with $\rho(t,\mathbf{x}')$, being the probability density to observe the
particle in point $\mathbf{x}'$ at time $t$, and $\mathbf{j}(t,\mathbf{x}')$, the corresponding flux
density, is well known:
\begin{equation}
 \label{eq:continuity}
 \frac{\partial \rho(t,\mathbf{x}')}{\partial t} + \boldsymbol{\nabla}\mathbf{j}(t,\mathbf{x}') = 0.
\end{equation}

Equation ~\eqref{eq:continuity} could rewritten as follows:
\begin{equation}
 \label{eq:continuity1}
 \frac{\partial}{\partial t} \int_{|\mathbf{x}'|\le|\mathbf{x}|}\!\!\!\!\!d\mathbf{x}'\rho(t,\mathbf{x}') = - 
\int_{S}d\mathbf{S}\;\mathbf{j}(t,\mathbf{x}'),
\end{equation}
where the integration is limited by a sphere $S$ of radius $|\mathbf{x}|$.  If the radius $|\mathbf{x}|$ is
sufficiently large and time $t$ is so small that the wave packet is not dispersed significantly (both conditions are
satisfied if $|\mathbf{x}|\gg\sigma_{\rm x}(t)$) then one might expect that the integral is almost saturated: 
\[
 \int_{|\mathbf{x}'|\le|\mathbf{x}|}d\mathbf{x}'\rho(t,\mathbf{x}') \approx 1
\]
and \[\frac{\partial}{\partial t}\int_{|\mathbf{x}'|\le|\mathbf{x}|}d\mathbf{x}'\rho(t,\mathbf{x}')\] vanishes.
Therefore, the right hand side of ~\eqref{eq:continuity1} also vanishes. This implies that the flux density should
decrease faster than $1/|\mathbf{x}|^2$. The flux density example for the non-covariant Gaussian wave packet in
~\eqref{eq:flux_non_covariant} agrees with this conclusion.

Let us perform the time integration for the left and right-hand sides between zero and infinity. Therefore, 
\begin{equation}
 \label{eq:continuity2}
 \int_{|\mathbf{x}'|\le|\mathbf{x}|}d\mathbf{x}'\rho(\infty,\mathbf{x}')-
\int_{|\mathbf{x}'|\le|\mathbf{x}|}d\mathbf{x}'\rho(0,\mathbf{x}')=  - 
\int_{S}d\mathbf{S}\;\boldsymbol{\Phi}(\mathbf{x}'),
\end{equation}
where $\boldsymbol{\Phi}(\mathbf{x}') = \int_0^\infty d t\; \mathbf{j}(t,\mathbf{x}')$ is the flux density
integrated over the time. By definition:
\begin{equation}
 \label{eq:prob_t}
 P(t,|\mathbf{x}|)\equiv \int_{|\mathbf{x}'|\le|\mathbf{x}|}d\mathbf{x}'\rho(t,\mathbf{x}')
\end{equation}
gives the probability to find a particle within a sphere of radius $|\mathbf{x}|$ at time $t$.
Apparently, due to dispersion of the wave packet, the probability to find a  particle within any volume of a finite
size tends to zero at $t\to \infty$ because the particle leaves the volume. On the other hand, at $|\mathbf{x}|$
much larger than the wave packet's ``size`` the value of $P(0,|\mathbf{x}|)$ is very close to unity because  initially
the wave packet is almost fully contained within a large enough volume. With these conditions
\eqref{eq:continuity2} becomes:
\begin{equation}
 \label{eq:continuity3}
 \int_{|\mathbf{x}|'\le|\mathbf{x}|}d\mathbf{x}'\rho(0,\mathbf{x}')=
\int_{S}d\mathbf{S}\boldsymbol{\Phi}(\mathbf{x}')\approx 1,
\end{equation}
from which it follows immediately that in the rest frame of the wave packet:
\[
 |\boldsymbol{\Phi}(|\mathbf{x}|)| = \frac{1}{4\pi |\mathbf{x}|^2}
\]
Therefore, the  $1/|\mathbf{x}|^2$ dependence holds true for any wave packet solution of the Klein-Gordon and/or Dirac
equations with finite normalization (whereas plain waves do not have a finite norm). At any given moment in time the
flux vanishes faster than  $1/|\mathbf{x}|^2$. 

\section{\label{sec:conclusion}Discussions and Conclusions}
Let us briefly discuss the main points raised in this paper. A wave packet possesses some characteristics of a
``particle`` or a solid body. It has well-defined mean energy and momentum while the wave packet is off-shell. In
other words, its energy in the rest frame of the wave packet is not equal to the mass of the waves composing the packet.
This can be easily seen from ~\eqref{eq:mean_momentum} in the rest frame:
\[
  \langle \text{E}\rangle  = \int\frac{d\mathbf{k}|\phi(\mathbf{k})|^2}{(2\pi)^32E_\mathbf{k}}E_\mathbf{k}\ge
m\int\frac{d\mathbf{k}|\phi(\mathbf{k})|^2}{(2\pi)^32E_\mathbf{k}} = m
\]
where we used the normalization ~\eqref{eq:rho_integral_3D}. 

A wave packet follows the classical trajectory for its mean
3-coordinate: $\langle\mathbf{x}\rangle=\langle\mathbf{v}\rangle t$. On the other hand, a particle described by the wave
packet can be detected at any place in the Universe, though the probabilities near the classical trajectory and far
away from it differ sharply.  

Wave packets, by their nature, disperse over time. The dispersion is very different between the direction along the
mean velocity and the transverse direction. We addressed this issue with help of Gaussian non-covariant and
covariant models, and found that the non-covariant wave packet predicts too slow a longitudinal dispersion relative to the covariant model. This makes a dramatic difference for ultra-relativistic particles like a neutrino with $\gamma=10^{10}$, which will not disperse significantly in the longitudinal direction during the lifetime of the Universe according to the non-covariant model.  However, using the covariant model, and assuming an initial wave packet spatial size on the order of 1 micrometer, such a neutrino will disperse in about $10^{-4}$ seconds.

By performing calculations for a wave packet of arbitrary form we confirmed in Sec.~\ref{sec:generalWP} that the relationship between the longitudinal and transverse dispersion times given by the covariant Gaussian model is correct. Moreover, we found a general formula for the dispersion of a wave packet of arbitrary form and found that it linearly increases with time with a coefficient of proportionality equal to the wave packet's velocity dispersion.

Finally, we addressed the question of interpretation of the dispersion of a wave packet, which is often considered as a disadvantage in attempts to describe a ``stable particle`` whose wave function vanishes with time. Proceeding from simple examples of Gaussian wave packets describing a spinless particle, generalizing it to wave packets of arbitrary form, and finally considering the continuity equation we find that the time-integrated  flux or probability density always displays an asymptotic behavior which is proportional to $1/|\mathbf{x}|^2$ in the rest frame of the wave packet, as one would expect for an ensemble of classical particles if its number density is normalized to the number of particles in the ensemble. The analysis of correspondence between an ensemble of one-particle wave packets and uniform flux of particles was also discussed in~\cite{Goldberger:1964,Feranchuk:2010}.

As we have demonstrated in this paper the origin of $1/|\mathbf{x}|^2$ law for quantum objects is their dispersion with time. For wave packets sufficiently narrow in time this law appears automatically provided the detection time is much longer than the wave packet time width. On the other hand accelerator experiments with beams very narrow in time or experiments at short enough distances might probe deviations from $1/|\mathbf{x}|^2$ law. 

It is my pleasure to thank E.Akhmedov, C.Kullenberg, S.E.Korenblit for useful discussions.

This work was supported by the Federal Target Program ``Scientific and scientific-pedagogical personnel of the innovative Russia'', contract No.14.U02.21.0913 as well as by RFBR grant 12-02-91171.

\end{document}